# Aitomia: Your Intelligent Assistant for AI-Driven Atomistic and Quantum Chemical Simulations

Jinming Hu,[1†] Hassan Nawaz,[1†] Yi-Fan Hou,[1†] Lijie Chi,[2] Xin-Yu Tong, Yuting Rui,[1] Yuxinxin Chen,[1,2*] Arif Ullah,[3*] and Pavlo O. Dral[1,2,4,5*]

[1] *State Key Laboratory of Physical Chemistry of Solid Surfaces, Department of Chemistry, College of Chemistry and Chemicals Engineering, and Fujian Provincial Key Laboratory of Theoretical and Computational Chemistry, Xiamen University, Xiamen 361005, China.*

[2] *Aitomistic, Shenzhen 518000, China.*

[3] *School of Physics, Anhui University, Hefei 230601, Anhui, China.*

[4] *Institute of Physics, Faculty of Physics, Astronomy, and Informatics, Nicolaus Copernicus University in Toruń, ul. Grudziądzka 5, 87-100 Toruń, Poland.*

[5] *Institute of Advanced Studies, Nicolaus Copernicus University in Toruń, ul. Wileńska 4, 87-100 Toruń, Poland.*

[†] *Co-first authors.*

*Email: chenyxx@stu.xmu.edu.cn (Y.C.), arif@ahu.edu.cn (A.U.); dral@xmu.edu.cn (P.O.D.)*

**Abstract**

We have developed Aitomia – a platform powered by AI to assist in performing AI-driven atomistic and quantum chemical (QC) simulations. This evolving intelligent assistant platform is equipped with chatbots and AI agents to help experts and guide non-experts in setting up and running atomistic simulations, analyzing simulation results, and summarizing them for the user in both textual and graphical forms. Aitomia combines LLM-based agents with the MLatom platform to support AI-driven atomistic simulations as well as conventional quantum-chemical calculations, including DFT, semiempirical methods such as GFN2-xTB, and selected high-level wavefunction-based methods, through interfaces to widely used programs such as Gaussian, ORCA, PySCF, and xtb, covering tasks from ground-state and excited-state calculations to geometry optimization, thermochemistry, and spectra simulations. The multi-agent implementation enables autonomous execution of complex computational workflows, such as reaction enthalpy calculations. Aitomia was the first intelligent assistant publicly launched on cloud computing platforms for broad-scope atomistic simulations (Aitomistic Lab@XMU at https://atom.xmu.edu.cn and Aitomistic Hub at https://aitomistic.xyz). Aitomia lowers the barrier to performing atomistic simulations, thereby democratizing simulations and accelerating research and development in relevant fields.

## Introduction

Computational chemistry has played a transformative role in advancing chemical research over the past decades, enabling spectra simulations and interpretation, the elucidation of reaction mechanisms, and the generation of training datasets for artificial intelligence (AI)-assisted design and discovery.[1-5] The backbone of such atomistic simulations is quantum chemistry (QC) based on first principles, solving the Schrödinger equation by employing numerous approximate methods that balance speed and accuracy. AI/machine learning (ML) methods are also increasingly used for AI-driven atomistic simulations. However, the growing sophistication of computational chemistry tools presents significant challenges for researchers.

Proficiency with these tools requires not only familiarity with complex underlying theories and package-specific options but, often, skills in Linux and access to high-performance computing (HPC) resources. Moreover, simulating realistic chemical processes often necessitates the integration of multiple computational packages, compounding the complexity for users across the chemical sciences. Despite progress in automating computational workflows,[6] two critical barriers remain. First, users must manually configure crucial simulation parameters, often requiring consultation of extensive and technical user manuals. Second, many workflows depend on HPC resources, which are not readily available to non-computational researchers, including experimental chemists and students.

Some of the aforementioned issues are alleviated with graphical user interfaces (GUIs) designed for the specific software packages. For example, a popular Gaussian program[7]for QC calculations includes GaussView as its official GUI, which facilitates the creation of input files and the analysis of calculation results. Many other packages, such as Materials Studio and ADF[8], were built to provide seamless experiences for computing job preparation, performing, and analyzing the computations. Unfortunately, GUIs do not fully solve the above problems, as researchers may still need to learn the GUI's options in addition to those of the underlying software, submit jobs to HPC clusters, and seek expert advice.

Another useful solution lies in the growing number of web-based platforms that provide simplified ways to submit and run calculations, often with web GUIs; examples include TeraChem[9] and ChemCompute.[10] We are also involved in the development of similar platforms, Aitomistic Hub[11] and Aitomistic Lab@XMU[12] (successor to the Xiamen Atomistic Computing Suite,[13] XACS), which enable QC- and AI-driven atomistic calculations on HPC clusters via a web browser. An interesting platform is Chemvox,[14] which enables researchers to run calculations using AI-powered speech recognition, allowing them to perform basic calculations via voice commands. Another attempt to bring QC calculations to general users, including students, is the mobile app MoleculAR, which converts 2D handwritten structures into 3D visualizations to improve understanding of molecular structures.[15] The platforms mentioned above often provide free HPC resources for basic use, such as education, and are convenient options for performing QC and ML simulations; however, users still need to understand the software's background and the supported theoretical methods. These website platforms also provide limited support for interpreting and analyzing the results.

With the success of large language models (LLMs) such as the GPT series [16-18] from OpenAI and Qwen[19], it is natural to ask whether LLMs can provide an ultimate solution for making QC- and ML-driven atomistic simulations accessible to a wider research community. Many studies have been combining LLMs with professional chemistry-related fields in recent years,[20-23] and, unsurprisingly, this trend also extends to atomistic simulations. One recent noteworthy attempt is the AutoSolvateWeb platform[24] for submitting jobs to simulate explicitly solvated molecules. This platform uses an AI agent that guides the user in selecting calculation parameters and provides explanations for basic queries; however, it is limited to submitting jobs to specific cloud computing services and provides neither analysis of the results nor any technical discussion of the calculations. This approach resembles a rule-based method to guide the users with rather limited explanations. The feasibility of using LLMs for autonomous simulations, from the design of calculations to the reporting of results and analyses, has been demonstrated in polymer research.[25] LLM assistant MDcrow was recently developed to automate molecular dynamics (MD) simulations.[26] A more general multi-agent system, El Agente, is powered by LLMs to generate and execute complex workflows, e.g., involving several ground-state single-point energy calculations, geometry optimizations, and frequency analysis using DFT and semi-empirical quantum-mechanical (QM) methods.[27] Efforts toward LLM-powered autonomous simulations are gaining momentum, as evidenced by the fact that the above works appeared online within five months, and the latter independent preprint on El Agente is dated a few days before the first version[28] of this preprint.

While the above research shows the great advancement of LLM accelerating the automation in QC field, at the time of the writing, until our work, there has been no online accessible platform providing an integrated, all-in-one, solution for both AI/ML and QC simulations powered by AI agents, and providing a comprehensive support of a wide range of typical simulation tasks ranging from ground- to excited-state single-point calculations, geometry optimizations, spectra simulations, etc. Consequently, there is a pressing need for a user-friendly platform that reduces technical and computational barriers, enabling a wider audience to effectively leverage computational chemistry tools.

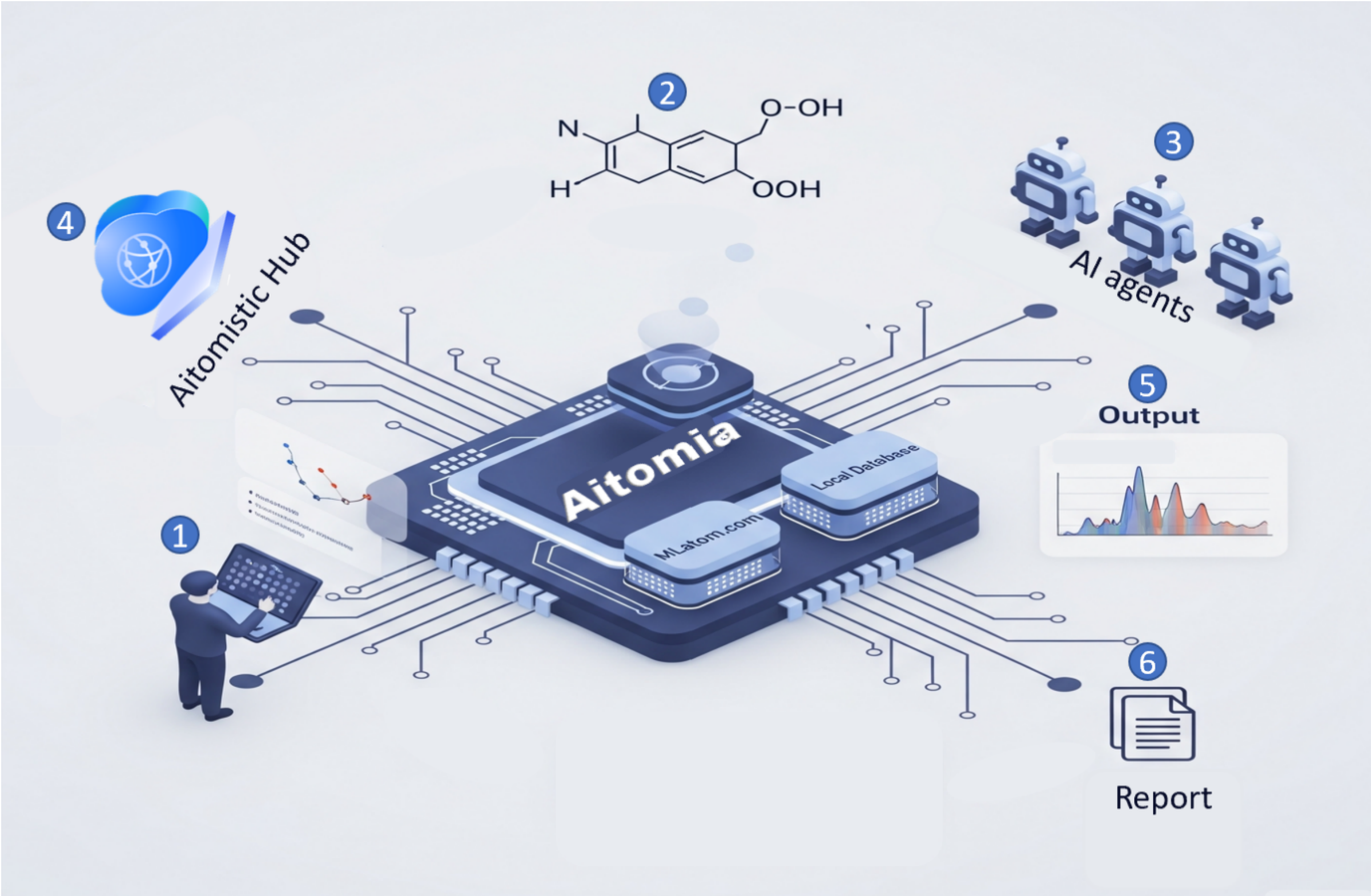

**Figure 1.** The overview of intelligent assistant Aitomia's place in AI-driven and QC atomistic simulations, with a chatbot interacting with the user, integrating the MLatom package for performing AI-enhanced computational chemistry calculations on the cloud computing platforms Aitomistic Lab@XMU[12] (https://atom.xmu.edu.cn) and Aitomistic Hub[11] (https://aitomistic.xyz), and providing final calculation results.

Here we present *Aitomia* – a platform powered by an AI assistant and agents to simplify the AI-driven atomistic and QC simulations (Figure 1). This platform is equipped with chatbots that communicate with users in natural language. The chatbots assist researchers at all stages, from setting up calculations and providing possible options to automatically performing simulations and analyzing their results in a clear, reusable format. Aitomia is leveraging the MLatom ecosystem[29] that supports a wide array of AI-driven atomistic and QC methods, as well as interfaces to various powerful electronic-structure and ML programs such as PySCF, Gaussian, Orca, xtb, TorchANI, etc., enabling essential computations such as energy calculations, geometry optimizations, thermochemistry, reaction, and spectra simulations.

Aitomia is a rapidly evolving system deployed on cloud computing platforms, such as Aitomistic Lab@XMU[12] (https://atom.xmu.edu.cn, free for academic users) and Aitomistic Hub[11] (https://aitomistic.xyz, free registration). Deployment of Aitomia on cloud computing platforms further democratizes access to computational chemistry. It augments Aitomia with a convenient GUI for molecular visualization, browsing, and analysis of calculation results, job submission, and other tools for atomistic simulations. Our agentic platform enables experts and non-experts alike, experimental researchers, educators, and students to utilize advanced computational chemistry tools, expanding their accessibility and applicability, ultimately accelerating advancements in the relevant fields, such as drug and materials design.

## Results and discussion

### *The scope of atomistic simulations*

Our goal is to design an intelligent assistant capable of performing most key computational chemistry tasks using a wide variety of state-of-the-art AI atomistic models and conventional QC methods. This is an ambitious goal that presents a formidable challenge, as it requires supporting many different software packages and theoretical methods, both QC- and AI-driven. Luckily, we can leverage the existing MLatom software ecosystem,[29] which is particularly well-suited to this goal because it seamlessly supports both QC and AI-driven atomistic simulations. Hence, our intelligent assistant, Aitomia, achieves its goal by leveraging MLatom's wide range of features. The current capabilities of Aitomia are summarized in Figure 2.

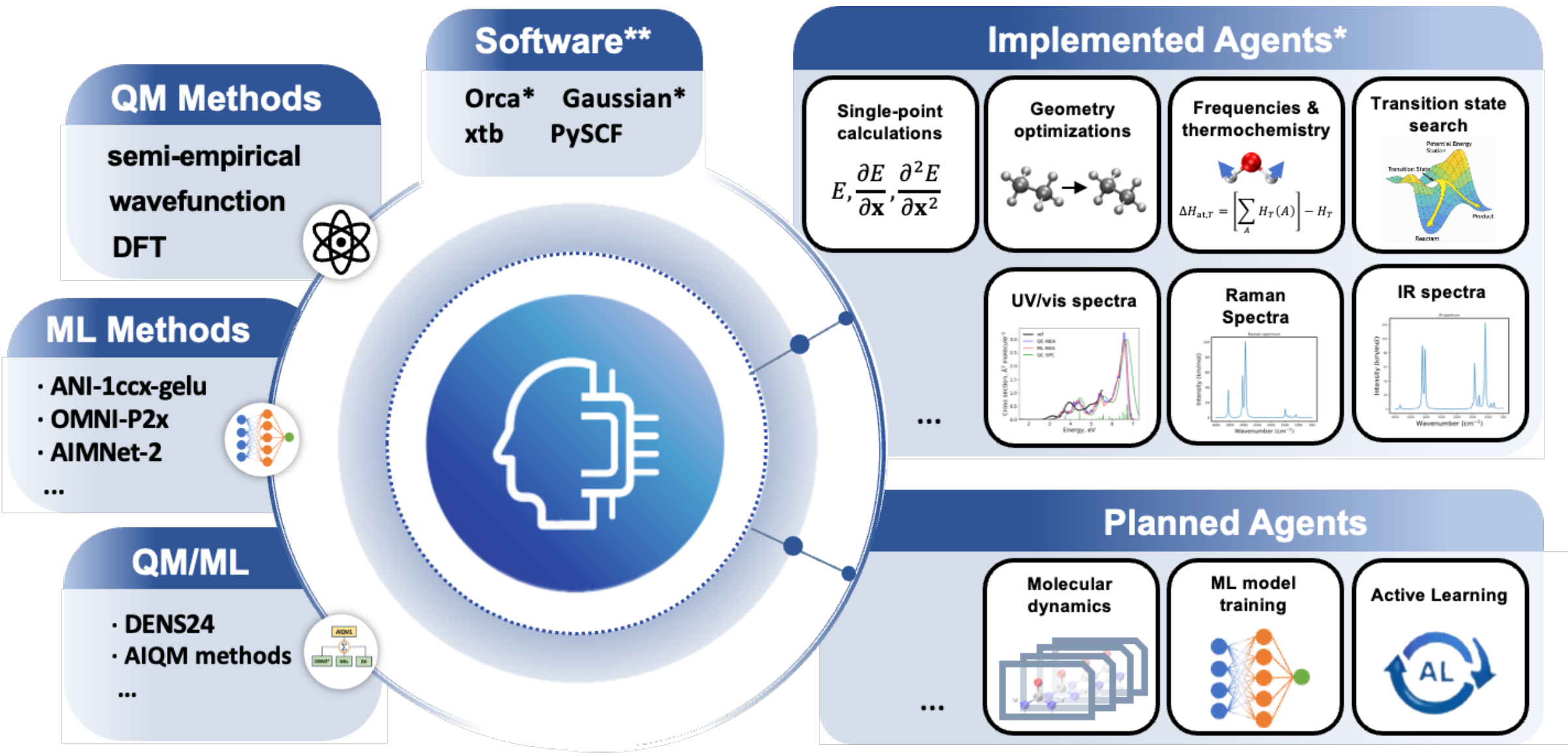

**Implemented agents can be used in more complex workflows, such as reaction enthalpy calculations.*
***Only software with permissive licenses is available online for registered users, while Aitomia supports more programs.*

**Figure 2.** Aitomia can perform a wide range of computational chemistry tasks using both AI/ML atomistic models and conventional quantum-chemical methods, including DFT, semiempirical approaches such as GFN2-xTB, and selected higher-level wavefunction-based calculations accessed through MLatom interfaces to external programs such as Gaussian, ORCA, PySCF, and xtb. Note that only software with permissive licenses is available online for registered users. The figure shows the single-task agents implemented in Aitomia, e.g., for geometry optimizations and thermochemistry, which can be combined into more complex workflows, such as reaction enthalpy calculations. A selection of the planned agents is also shown. Some of the figure elements are adapted from J. Chem. Theory Comput. 2024, 20, 1193–1213 under the CC-BY license.

Currently, Aitomia can perform tasks such as single-point calculations, geometry optimizations, frequency and thermochemistry calculations, and IR and UV/vis spectra simulations. For this, it leverages many state-of-the-art pre-trained universal ML models, such as the AIQM,[30-33] ANI,[34-37] OMNI-P,[38, 39] and AIMNet[40] series. Some of these models, such as the AIQM series and ANI-1ccx, approach the coupled-cluster accuracy at a fraction of the computational cost of density functional theory (DFT) for many tasks. OMNI-P2[39] is a unique universal potential for excited-state calculations, furnishing UV/vis absorption spectra with accuracy approaching that of TD-DFT. We are planning to extend this functionality to other tasks supported by MLatom, e.g., molecular dynamics, training, evaluating, and using custom ML models, building data sets with active learning, and fine-tuning the universal ML models for user-specific tasks. Aitomia can also combine the single tasks into more complex workflows, such as the calculation of the reaction enthalpies, as we will discuss separately.

***Aitomia as an assistant for performing standard computational tasks***

Typically, performing single computational tasks, such as geometry optimizations or frequency calculations, requires users to be familiar with the corresponding software input formats and options. Learning them is a labor-intensive process, impeding the widespread adoption of various software and new techniques. To overcome this barrier, we have built AI agents for Aitomia, which invoke LLMs to interact with the user, choose an appropriate computational task requested by the user, submit the computing job to the HPC server, retrieve the output, analyze it, and report the results back to the user in text and graphical format. Additionally, we implemented quality-of-life features that enable Aitomia to retrieve molecular structures from user-provided chemical names. With Aitomia's assistance, AI-driven atomistic and QC calculations can be executed and analyzed autonomously, with the input and output files and scripts organized in a clear directory structure for the user's review and results retrieval.

We show an example in Figure 3. The user interacts with Aitomia and requests the structure of the hexanol molecule, which Aitomia returns, allowing the user to analyze it in the web GUI (Figure 3a). Later, the user requests infrared (IR) spectra calculations for this molecule, and Aitomia calculates them using the AIQM2 method[33] (Figure 3b). The calculations finish in seconds, and the resulting spectrum is very close to the experimental spectrum retrieved from the NIST website[41] (Figure 3c). Importantly, calculations of similar quality would take much longer with DFT, whereas analogous spectra simulations with semi-empirical QM methods would not be as accurate.[42] Interactions can be performed in many other languages, such as English, Chinese, and Ukrainian, which we tested. Videos of this and other examples are available on the Aitomia website at http://mlatom.com/aitomia.

We note that Aitomia prefers using the AIQM methods (e.g., the choice of AIQM2 in the above IR calculation), because they are among a few methods that allow fast, accurate, and robust simulations for a broad range of systems, while other approaches such as DFT are often too slow (particularly for the free online platform), or too inaccurate (e.g., semi-empirical methods such as GFN2-xTB and others, which are less accurate than AIQM series according to numerous benchmarks)[30-33, 43, 44], or not robust and transferable enough (e.g., more recent universal machine learning potentials which often break down in uncommon situations).[43]

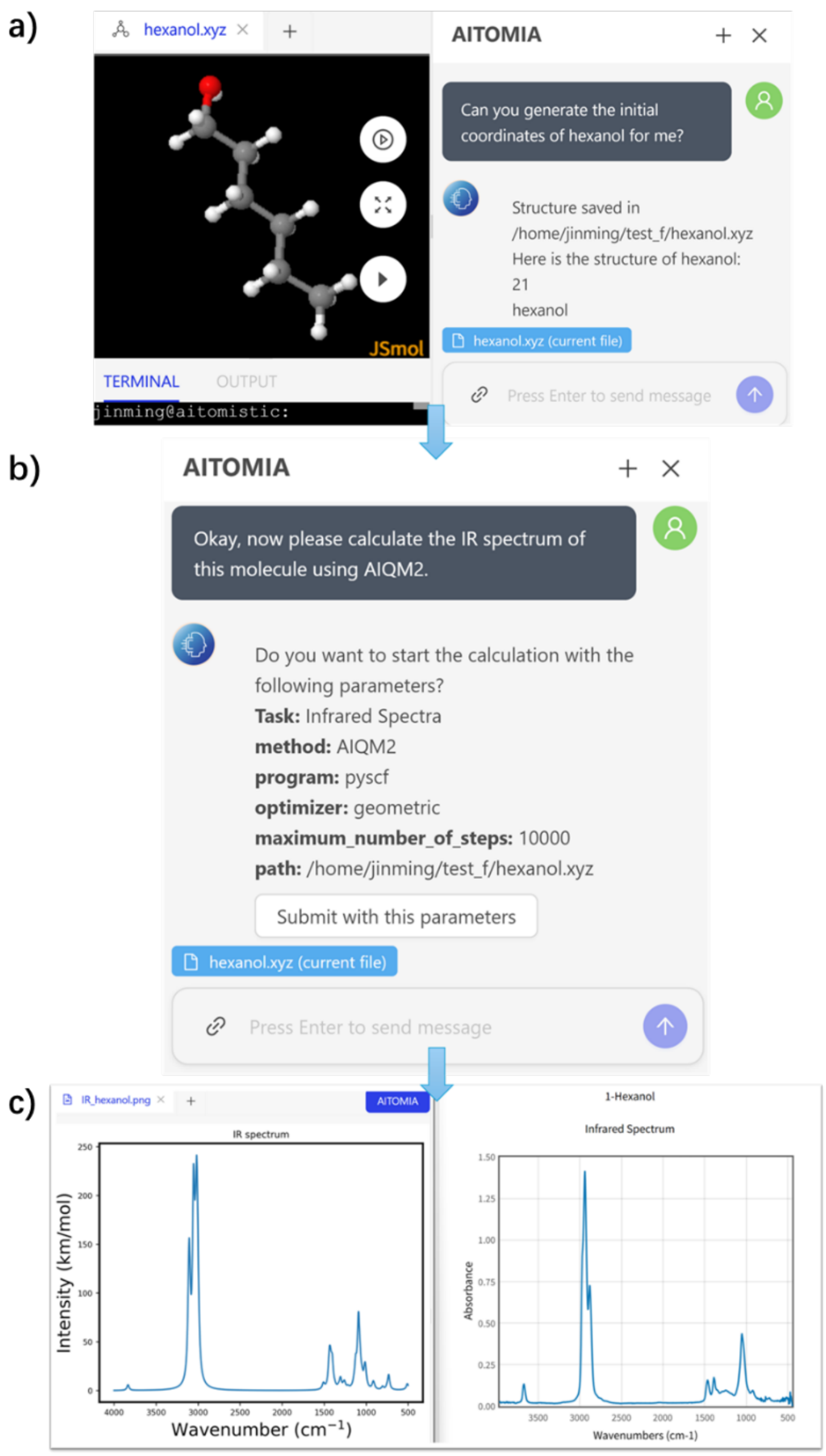

**Figure 3.** An example of performing infrared spectra calculations with Aitomia on the cloud computing platforms, guiding the user through a) structure generation, b) calculations, and c) results. The videos of this and other examples are available at http://mlatom.com/aitomia. The experimental spectrum is taken as a screenshot from the NIST web page[41]. Note that all screenshots were taken in January–February 2026 and are provided here for reference only, while the actual behavior is subject to change as the platform is continuously upgraded.

Hence, Aitomia simplifies general, frequently used computational chemistry tasks by allowing users to perform them via natural-language chat, eliminating the need to manually learn and write input files for various software packages. However, the researchers often need to perform multiple such computations, e.g., to calculate reaction and isomerization energies, binding energies, electron affinities, etc. This necessitates the development of a more autonomous assistant capable of handling such complex workflows automatically.

***Design of autonomous workflow and execution with multi-agent Aitomia***

Beyond simplifying typical single computational tasks, the goal of Aitomia is to assist with performing full computational workflows comprising multiple tasks. We enable these workflows via an autonomous, multi-agent extension of Aitomia. It is capable of designing workflows and executing them with minimal manual intervention.

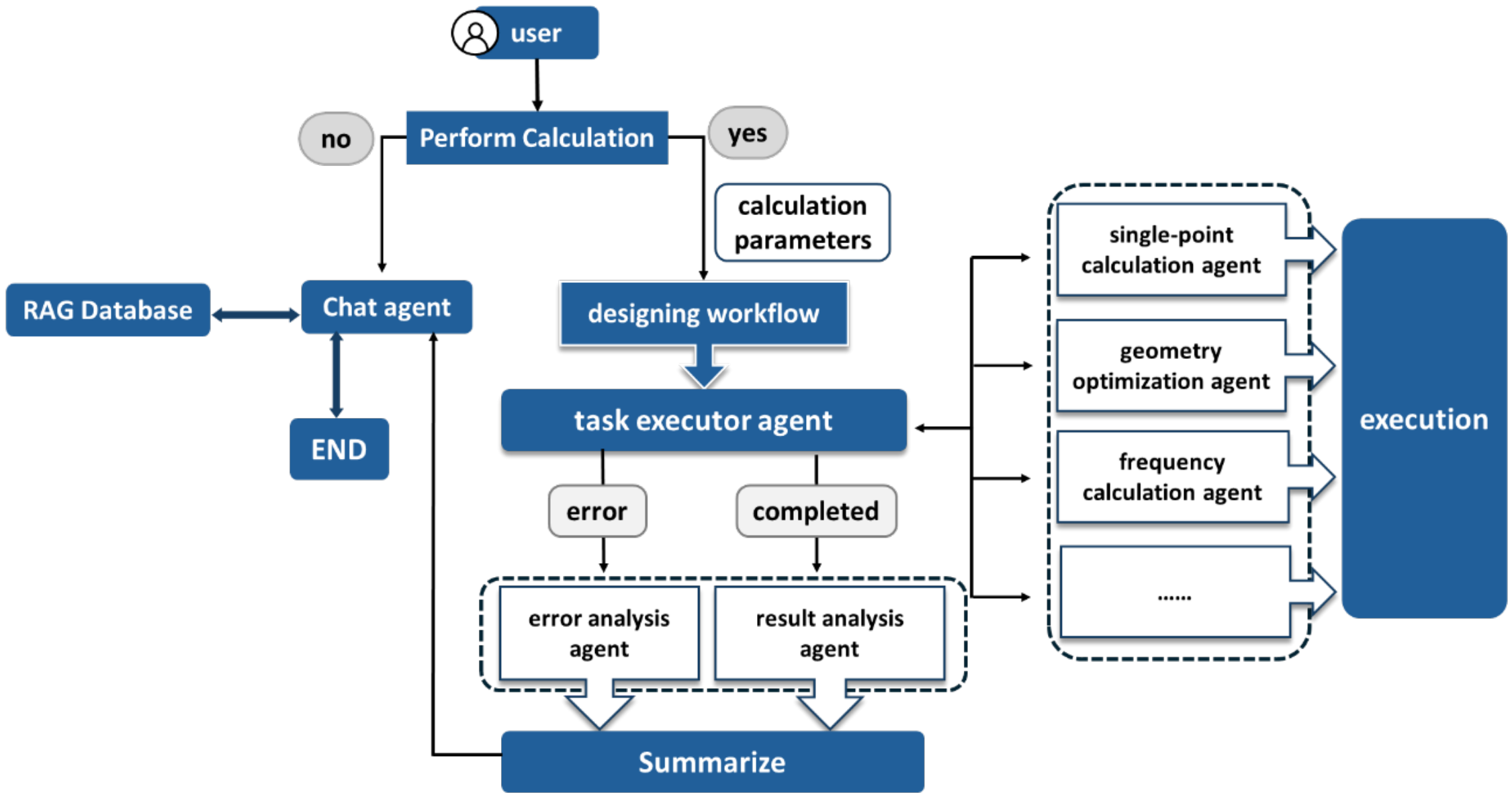

**Figure 4.** The architecture of multi-agent Aitomia enabling complex computational workflows.

First, the multi-agent system evaluates the user's intent to determine whether a computational task is required. If computation is needed, the system validates the relevant computational parameters and, under the LangGraph-based orchestration layer's control, activates the executor agent. The executor agent initializes the working directory, stores the user-provided initial molecular structure (if available), constructs the overall computational workflow, and dispatches subtasks to subordinate agents.

Each sub-agent acquires the necessary contextual information, including the working directory. If the user does not provide an initial molecular structure, the sub-agent invokes a structure-retrieval agent to obtain the required molecular geometry. Within the structure-retrieval agent, the molecular species required for the current task are first identified, after which the agent determines whether the corresponding structure should be retrieved from external sources or reused from previously generated results. The retrieved structure is then validated to ensure its chemical plausibility before being returned to the calling sub-agent.

For sub-agents responsible for simpler tasks, they directly call MLatom-based scripts to perform the required calculations. In contrast, sub-agents handling more complex tasks further refine the computational workflow and may invoke additional specialized agents to decompose the task and execute specific components.

After completing all assigned computations, each sub-agent performs local result analysis and returns the processed outputs to the executor agent. Once all sub-agent responses are collected, the executor agent conducts an integrated analysis, summarizes the key outcomes of the computation, records the essential information in the system state, and terminates the current computational workflow. The entire process concludes with a report to the user summarizing either the results or encountered errors.

We will use the reaction energy for the Diels–Alder reaction of cyclopentadiene and maleimide as a demonstration of the application of the multi-agent Aitomia. We only present the main points of the chat between the user and Aitomia in Figure 5; complete model answers and other examples are provided in the Supporting Information (SI). In this example, Aitomia completes the task in under 7 minutes and, overall, calls single-task agents 12 times. The final reaction energy of −33.7 kcal/mol is rather close to the best theoretical estimate of ΔE = −34.2 kcal/mol in the GMTKN55 database[45], due to the use of an accurate AIQM2 method, which is orders of magnitude faster than DFT. Aitomia also provides other thermochemical properties for this reaction, i.e., enthalpy and Gibbs free energy changes.

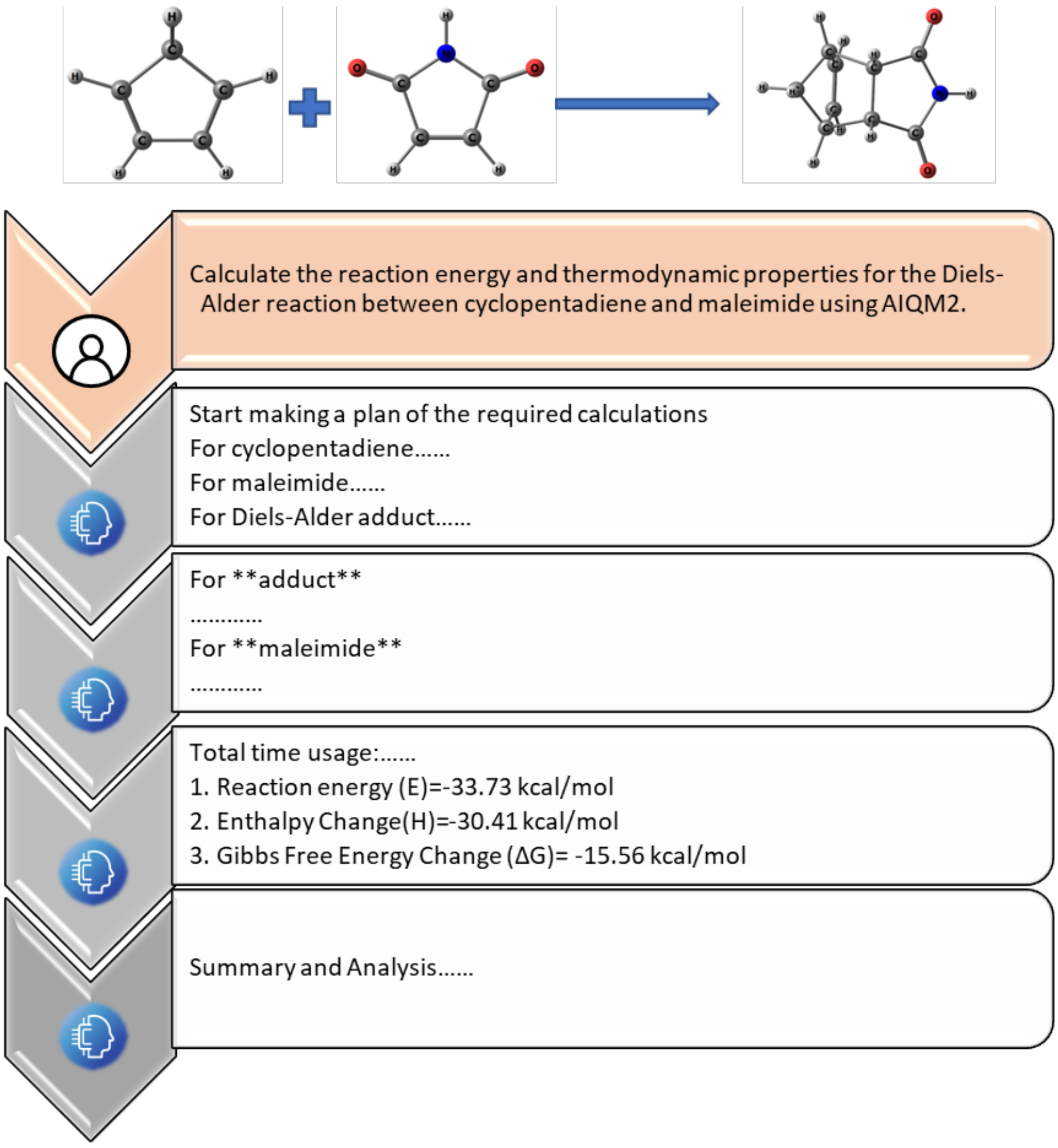

**Figure 5.** The use of Aitomia's multi-agent implementation for the autonomous calculation of reaction energy and thermodynamic properties for the Diels–Alder reaction between cyclopentadiene and maleimide.

We provide additional examples in the Supporting Information (Examples F–T, Pages S7–S86), which, taken together, demonstrate Aitomia's ability to handle complex workflows.

We note that the multi-agent system, e.g., the use of multi-agent Aitomia, requires extensive LLM-based reasoning and, hence, is slower and more resource-intensive than using Aitomia for single-task computations. Also, the autonomous use of Aitomia for increasingly complex workflows is more prone to occasional hallucinations and mistakes. For example, in our initial tests, we observed that when running the same prompt for the Diels–Alder reaction (see example above), it sometimes failed to properly convert hartree to kcal/mol (the value in hartree was still correct). We fixed this unit-conversion problem; in general, the user is advised to critically evaluate the responses from the multi-agent Aitomia. More benchmarks are provided next.

***Benchmark on atomistic simulations tasks***

A major challenge for Aitomia is correctly interpreting user queries, determining which information is missing and which calculations are required, and finally organizing the corresponding computational procedures to obtain reliable results. Thus, we conducted a systematic benchmark of Aitomia on a set of representative theoretical chemistry tasks, assigning each task a difficulty level based on the degree of autonomy required to answer the given questions. The full overview of the tasks is provided in the SI benchmark spreadsheet. Tasks with low autonomy typically provide clear instructions regarding the type of calculations needed and involve relatively simple execution with minimal orchestration. Medium-level autonomy tasks require a higher level of routing to organize multiple computational steps, but still contain explicit instructions describing the necessary procedures. In terms of high-level autonomy, little information is provided to decide which tasks to perform, and the queries are much closer to what we as humans face every day. It requires a deep understanding of the user's questions and the correct calculation procedure to obtain the results. We used low and part of medium-level autonomy to implement Aitomia; hence, they are marked 'seen' in the SI benchmark spreadsheet, while all the others were not seen by developers and are marked accordingly.

A task is considered successfully completed if it terminates normally and the deviation between the results generated by Aitomia and those obtained from manual calculations is sufficiently small. Specifically, for tasks requiring a precise numerical answer, the energy deviation must be less than $10^{-6}$ hartree, and deviations for other properties (geometries, frequencies, etc.) must be below $10^{-4}$. In addition, the final analysis provided by the system must reasonably address the user's query and avoid significant chemical inaccuracies, which we also assess. To assess robustness and reproducibility, each benchmark task was independently executed 10 times under identical settings. The results demonstrate consistent performance across repeated trials, indicating stable agent orchestration and reliable task execution. The specific task types, representative examples for each category, and the corresponding benchmark results are summarized in Table 1. The full task descriptions, their corresponding abbreviations, and the success rates across independent runs are provided in the SI spreadsheet. Because Aitomia is now actively being developed with improved intelligence, the benchmark results reflect only its performance using versions from January to February 2026.

**Table 1.** Classification of benchmark tasks and representative examples across different levels of difficulty. Abbreviations for each task are provided in parentheses.

| Level of Autonomy | Description | Examples | Number of Examples | Avg. Success Rate |
|---|---|---|---|---|
| Low | Clear instructions regarding the type of calculations are provided in query with simple execution of minimal orchestration | Calculate ground-state energy and gradients for acetone molecule using AIQM2.<br>Perform geometry optimization for p-xylene using AIQM1 method, use geometric as optimization program.<br>Calculate IR spectrum for acetonitrile using AIQM2 method, use geometric for optimization and PySCF for IR calculation.<br>Calculate UV-vis spectrum for naphthalene using semi-empirical method OM2, and do optimization using AIQM2 first, set band width 0.4 and contribution threshold 0.05. | 27 | 99.6% |
| Medium | Hints to type of tasks are given in the question. Questions have to be answered according to computational results. | Investigate how conjugation length affects UV absorption by comparing ethene, butadiene, and hexatriene. Analyze the red shift trend.<br>Which tautomer of acetaldehyde (keto vs enol) is more stable in the gas phase? Compare their thermodynamic properties and IR spectra.<br>Calculate the reaction energy and thermochemistry properties of Claisen condensation of two ethyl acetate using AIQM2 method<br>Calculate the proton affinity of ammonia ($NH_3$) using wB97X/Def2-SVP | 11 | 70.9% |
| High | The general chemical questions that require comprehension of the context and perform required calculations to get the answer. | Does 2-methoxytetrahydropyran obey classical steric expectations, or is its conformational equilibrium dominated by stereoelectronic effects?<br>Compare the UV absorption spectra of benzene and aniline to analyze the effect of amino substitution on electronic transitions.<br>Compare the intramolecularly hydrogen-bonded and non-hydrogen-bonded conformers of salicylic acid in | 9 | 45.6% |

| | | |
|---|---|---|
| | | the gas phase. Determine their relative thermodynamic stability and analyze how hydrogen bonding influences the O–H stretching frequency and IR intensity. |

It should be noted that, although challenging tasks demonstrate increased flexibility, they are fundamentally composed of basic agents. Consequently, complex tasks that rely on fundamental computational capabilities not supported by the benchmarked version of Aitomia—such as $pK_a$ calculations or solvation free energy evaluations that require unimplemented capabilities of the base agents—were beyond its scope.

In general, Aitomia achieves excellent stability on low-difficulty tasks, with a success rate approaching 100%. It provides a solid foundation for more complex tasks composed of these basic operations. For medium-difficulty tasks, which involve more complex workflows but still provide explicit instructions, Aitomia achieves a success rate of 70.9%. In contrast, the performance on high-difficulty tasks drops to 45.6%. An examination of failed or partially successful cases indicates that several systematic factors contribute to performance degradation.

One important factor is the absence of an automated self-correction mechanism in the current implementation. The present framework follows a relatively straightforward execution pipeline in which intermediate computational results are not actively validated. Consequently, errors introduced in earlier stages can propagate to the final output. For example, if a geometry optimization yields a structure with imaginary frequencies, the workflow proceeds without triggering reoptimization or structural correction, thereby compromising the reliability of subsequent thermodynamic or vibrational analyses. This behavior can be observed in several benchmark cases (e.g., MED.EWG, and HIGH.Gas.Vibrational, see the benchmarks in the SI for definitions and prompts). Notably, inspection of the generated reasoning indicates that the system often identifies the underlying source of the error in its final analysis. This observation indicates that integrating a self-correction module capable of diagnosing common computational issues and invoking specialized correction agents (e.g., imaginary-frequency removal or workflow redesign) could substantially improve the system's robustness.

A second source of error arises from the molecular retrieval stage, where the system occasionally fails to distinguish between isomeric and identical structures correctly. When the user-provided molecular description is ambiguous, the system may incorrectly interpret identical species as different isomers or, conversely, confuse distinct isomers as the same structure. This issue is particularly evident in cases such as HIGH.StericVsStereoelectronic. Even when the molecular identity is described relatively clearly, occasional misassignments still occur (e.g., MED.AxPref.StereoElec), suggesting that the current molecular identification procedure lacks sufficiently rigorous structural validation. In addition, this problem affects cases involving different conformational states of the same molecule (e.g., HIGH.CONF.HB-vs-nHB). When the user does not explicitly provide a molecular structure, the system retrieves molecules from a database based primarily on the molecular name.

However, such name-based retrieval typically returns only a single reference geometry and does not account for alternative conformations. Consequently, when tasks require comparing structurally distinct conformers—such as intramolecularly hydrogen-bonded versus non-hydrogen-bonded conformers—the system may inadvertently use the same retrieved structure for both calculations. This results in identical optimized geometries and energies, precluding a meaningful comparison.

In several benchmark tasks, the quantum chemical calculations themselves complete successfully, yet inaccuracies appear in the AI-generated chemical interpretation. These issues mainly arise in the post-analysis stage and typically involve minor errors in spectral assignments or mechanistic interpretation. For instance, in one of the benchmark runs (RUN7) of the HIGH.AminoSubstitution test case, small inaccuracies appear in the assignment of spectral features, while in (RUN10) of the same test, the electronic transition assignments for benzene were reversed. Although these issues do not affect the underlying numerical calculations, they reduce the reliability of the generated chemical explanations.

Another issue arises when the reaction equation is not explicitly specified by the user. In some cases (e.g., MED.ΔGdeprot), the LLM may infer an incomplete reaction scheme, such as omitting the $H^+$ species in deprotonation calculations.

Finally, a small number of failures are attributable to infrastructure-related issues, such as instability of the large language model service or unexpected server interruptions during task execution. While these factors are external to the computational framework itself, their impact could be mitigated through simple fault-tolerant mechanisms such as automatic task retry.

Taken together, these results suggest that Aitomia performs robustly on well-defined computational chemistry tasks, while performance decreases in scenarios involving ambiguous problem definitions, complex workflows, or automatic reaction inference. The benchmark analysis highlights several promising directions for future development, including incorporating automated self-correction mechanisms, improving molecular validation procedures, and adopting more robust reasoning strategies.

***Aitomia serves as a look-up tool for atomistic calculations***

Even with a powerful, intelligent assistant at hand, to ensure scientifically reliable results, users themselves must have background knowledge of simulations, how they are performed, and which QC methods or AI models are appropriate for the task. Before the rise of LLMs, the only ways were either learning from manuals and tutorials or from a human teacher. This severely limits the speed of knowledge acquisition, which might also be limited in access. It is also inconvenient to look up the information when needed in the lengthy manuals.

The challenge here is to create an LLM-driven lookup tool that provides reliable information based on the domain knowledge about the underlying MLatom capabilities exploited by Aitomia. We solve this challenge by exploiting the Retrieval-Augmented Generation (RAG) technique to improve the LLM's robustness by providing access to external knowledge (MLatom documentation), thereby reducing hallucinations and enhancing factual accuracy.

Aitomia, with access to the RAG database, provides a useful chatbot expert in atomistic simulations that can assist non-experts in providing background information on available computational approaches. Here, we show an example of asking Aitomia about the difference between two MLatom computational methods (Figure 6).

What is the difference between AIQM1 and AIQM2 in MLatom? Please make a table.

**Aitomia with RAG**

| Feature | AIQM1 | AIQM2 |
| --- | --- | --- |
| Generation | First-generation AI-enhanced QM method | Second-generation AI-enhanced QM method |
| Target Accuracy | Approaches CCSD(T) (gold-standard) accuracy | Targets chemical accuracy for organic reactions, often beyond DFT |
| Speed | Speed of semiempirical QM methods | High speed, competitive with or faster than DFT for reaction simulations |
| Applicability | Ground-state, closed-shell; also transferable to charged, radicals, excited states | Optimized for transition states (TS), reactive dynamics, organic reactions |
| Excited States | ✅ Supported | Not explicitly mentioned — likely less emphasis compared to AIQM1 |
| Heats of Formation | ✅ Calibrated and accurate | Not mentioned — likely not a focus |
| Local Installation | ✅ Possible | Not specified — assumed possible via MLatom |
| Cloud Stability (XACS) | ❌ Many numerical problems due to lack of analytical derivatives | ✅ More stable and much faster on cloud (implied by context) |
| Dispersion Correction | Includes explicit D4 correction for ωB97X | Not specified — may be included or handled differently |
| Underlying QM Method | Uses modified ODM2* (via MNDO or Sparrow) | Not explicitly stated — likely improved/updated underlying model |
| Molecular Properties | Dipole moments, charges, oscillator strengths, nonadiabatic couplings | Also provides dipole moments (e.g., for IR spectra), likely extended property support |
| Element Support | Limited to CHNO | Not explicitly stated — likely still CHNO or slightly extended |
| Recommended Use Case | General-purpose, calibrated thermochemistry, excited states | Organic reaction simulations, TS optimization, frequency calc, IR spectra, reactive dynamics |
| Citation (Primary Paper) | Zheng et al., *Nat. Commun.* 2021, *12*, 7022 | Chen & Dral, *Chem. Sci.* 2025, accepted manuscript |
| Successor / Alternative | Being succeeded by UAIQM (universal model) | Part of next-gen AIQM series; more robust for dynamics than AIQM1 |

**Aitomia without RAG**

| Feature | AIQM1 | AIQM2 |
| --- | --- | --- |
| Exists? | ✅ Yes — Officially published and implemented | ❌ No — Not published or implemented |
| Description | Hybrid AI-QM method combining DFT, ML, and semiempirical corrections for accurate energies | Does not exist; may be confused with future work or misinterpretation |
| Reference | *Nat. Commun.* 13, 5559 (2022) | — |
| Available in MLatom? | ✅ Yes | ❌ No |
| Purpose | High accuracy at low cost for ground-state energies | — |
| Components | ωB97X/6-31G*, D4 dispersion, dHNN, ODM2* | — |
| Developers | Pavlo O. Dral et al. | — |

**Figure 6.** Aitomia can provide non-experts with background information on the details of the platform's underlying methods. Here is an example showing Aitomia's response to an MLatom-related question with and without RAG.

In the example, Aitomia has been asked, "What is the difference between AIQM1 and AIQM2 in MLatom?" Aitomia satisfactorily answered this question using RAG across various aspects, including accuracy, speed, and applicability. The information is mostly correct except for the supported element of AIQM2, which is explicitly stated in the MLatom documentation. In contrast, Aitomia without RAG provided no information on AIQM2 because the LLM used only had knowledge as of 2024 June.

Overall, RAG enables Aitomia to answer questions in niche fields. It also has the potential to be employed as a teaching assistant in computational chemistry courses and to be incorporated into task agents to enhance their performance, which we plan to explore in the future.

### *Limitations*

Despite Aitomia's demonstrated flexibility, the presented capabilities were inherently constrained by the scope of the integrated fundamental functional agents. At that stage, Aitomia primarily focused on establishing a robust foundation for complex AI-driven atomistic and quantum chemical workflows, rather than exhaustively covering all possible simulation tasks.

Aitomia operated by decomposing complex scientific problems into a set of basic computational tasks. For example, reaction enthalpy calculations were automatically formulated as combinations of geometry optimization and frequency calculations for multiple molecular systems, with the overall task structure orchestrated by LLMs. While this design allowed for a high degree of generality and adaptability, the range of problems it could address remained limited by the availability, maturity, and reliability of these basic task modules.

Furthermore, the implementation supported only a limited set of molecular input formats, primarily XYZ and JSON representations. This restriction limited interoperability with existing simulation pipelines and user workflows, particularly when alternative or more specialized file formats were required.

Additionally, the integration of Retrieval-Augmented Generation (RAG) focused primarily on answering MLatom-related queries, whereas the calculation agents underused it.

## Conclusions

We have introduced an intelligent AI assistant, Aitomia, that helps researchers perform atomistic simulations tailored to their needs, even without specialized knowledge of the underlying software. It draws on the versatility of the MLatom software ecosystem to enable a broad range of computational chemistry tasks, spanning both traditional QM (DFT, HF, and post-HF; semi-empirical methods) and AI atomistic models (universal ML interatomic potentials and beyond).

Aitomia can assist with typical tasks such as geometry optimizations of minima and transition states, frequency and thermochemistry calculations, and simulations of IR and UV/vis spectra. Because of the extensive use of AI atomistic models, calculations are often much faster than DFT, with better or comparable accuracy. In addition, Aitomia can autonomously design and execute complex computational workflows, enabling convenient calculations of reaction enthalpies, atomization energies, proton affinities, and related quantities using single-task AI agents. Aitomia can also be used as an expert advisor for performing atomistic simulations. Its ability to analyze calculation results and answer common QC and ML questions plays a promotional role, helping professionals and non-professionals conduct computational research and analysis in related fields such as drug and materials design.

Aitomia is the first AI assistant of this kind, which is publicly accessible on a cloud computing service with accompanying GUIs, browsing, and analysis tools: it is deployed on the Aitomistic Lab@XMU[12] (https://atom.xmu.edu.cn, free for academic users) and Aitomistic Hub[11] (aitomistic.xyz, free registration), further contributing to the democratization of the atomistic simulations.

## Outlook

Aitomia is a rapidly evolving system, and updated versions supporting more complex, more autonomous end-to-end workflows are on the way, e.g., ML model training and active learning (Figure 7). In the future, as more advanced large language models and agentic workflow tools become available, they can be integrated to further enhance Aitomia's performance and robustness.

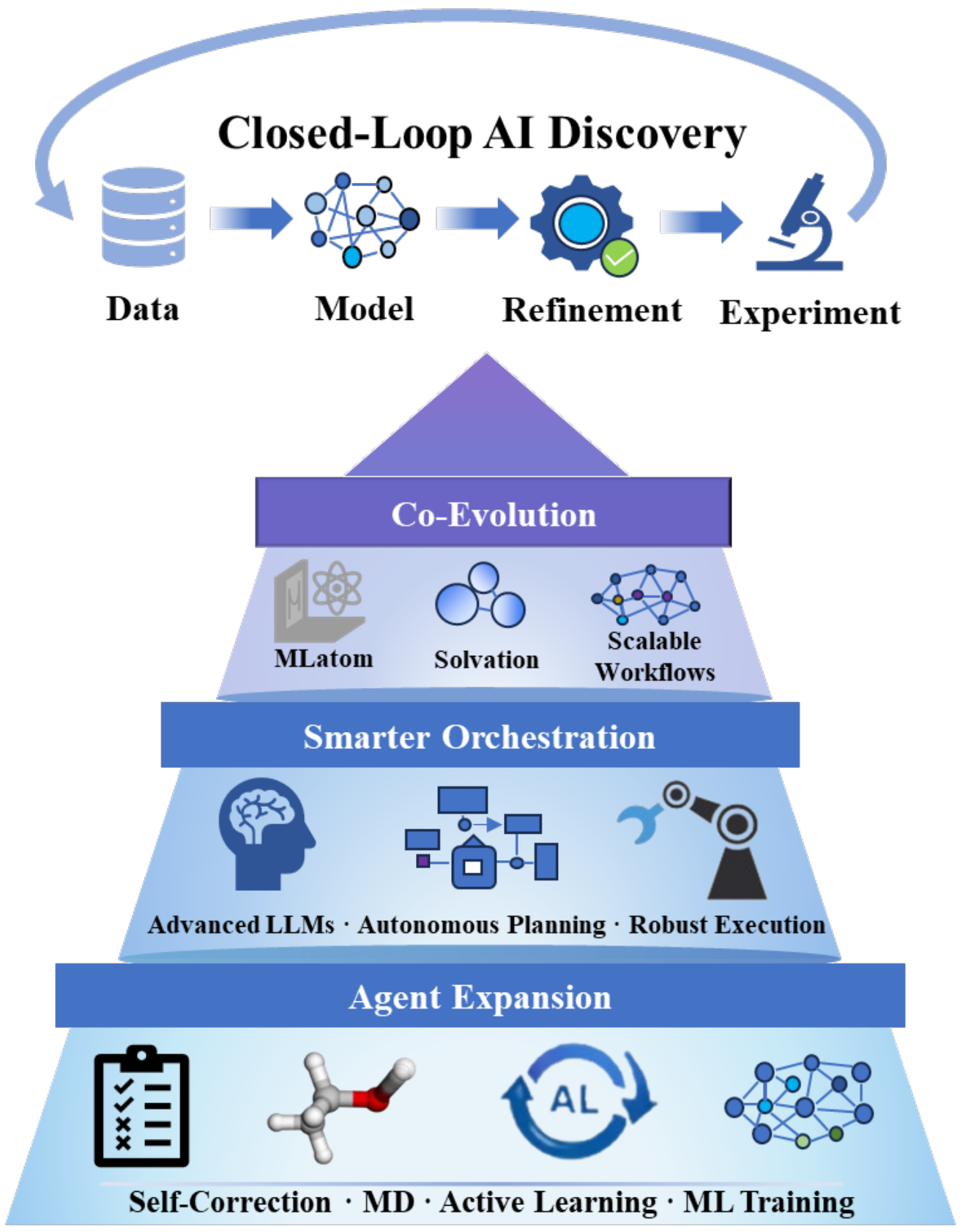

**Figure 7.** Aitomia development outlook.

In addition, an important direction for future development is the introduction of self-correction mechanisms, motivated by benchmarking analyses showing that a substantial fraction of current errors stems from the lack of automated error detection and recovery. In practice, many failures arise from common issues in computational chemistry workflows. For example, geometry optimizations may converge to saddle points rather than true minima, as indicated by the presence of imaginary frequencies in subsequent frequency calculations. In such cases, a self-correction mechanism could automatically trigger additional optimization steps, adjust computational methods or optimizers, or regenerate improved initial guesses to obtain physically meaningful structures. By enabling automatic validation of intermediate results and adaptive recovery from both computational and system-level errors, self-correction mechanisms are expected to substantially improve the stability, robustness, and reliability of autonomous computational workflows.

Future development of Aitomia will also focus on expanding the integration of independent functional agents for computational tasks supported by MLatom. In particular, incorporating additional capabilities, such as molecular dynamics simulations and broader support for molecular input formats, is expected to significantly expand the complexity and diversity of workflows that Aitomia can autonomously construct.

In parallel, further development of MLatom is planned in response to both user requirements and Aitomia's evolving needs. This includes incorporating solvation effects and developing machine-learning models specifically designed for solvated systems. The co-evolution of Aitomia and MLatom is inherently synergistic: the extensibility of MLatom enables continuous expansion of Aitomia's functionality, while the demands of Aitomia provide practical motivation for advancing MLatom.

Additionally, the integration of RAG will continue to evolve, allowing Aitomia to autonomously retrieve and incorporate information from external databases, scientific literature, and user-shared results. This will enable a more dynamic and adaptive system, capable of improving its responses and decision-making processes in real time as new data becomes available.

Beyond functional expansion, we aim to improve the overall framework's flexibility, motivated by advances in agentic workflows. Future versions of Aitomia will include more autonomous execution of increasingly complex tasks. In the longer term, we envision a closed-loop framework in which Aitomia can automatically collect computational and experimental data, train machine-learning models, and ultimately lead to iterative refining of these models. When combined with experimental automation platforms, this approach can enable continuous feedback between simulation, experimentation, and model improvement. This feedback loop would not only improve the performance of Aitomia's workflows but also enhance the reliability and accuracy of the underlying models over time.

## Methodology

### *LangGraph-based multi-agent architecture*

In the implementation of Aitomia, single computational tasks rely on the native tool-calling capabilities of large language models to parse user-provided parameters and invoke the corresponding computational functions. This version uses a lightweight, robust logic with relatively low latency in language-level parsing; consequently, the primary computational cost arises from the calculations themselves. However, this design offers limited flexibility and requires users to manually guide the workflow step by step to complete the desired computations.

The Atomia multi-agent architecture is implemented using LangGraph, and the overall structure of task execution is illustrated in Figure 4. For the agent system, we adopt a graph-controlled dynamic routing mechanism. Specifically, a dedicated decision agent makes a judgment on which agent should be called next by using an LLM tool to output a structured decision. This decision is then returned to LangGraph, allowing the graph engine itself to perform the transition.

In this design, the LLM participates in dynamic decision-making, but the execution flow remains explicitly governed by the StateGraph, ensuring reproducibility and clear separation between reasoning and control logic. The agents communicate through graph nodes, with the system featuring a main orchestrating agent responsible for analyzing user queries, designing the workflow, and orchestrating execution by calling upon corresponding single-task agents for specific calculations and file management. Each single-task agent is dedicated to an independent computational module (e.g., geometry optimization or reaction simulations) and possesses its own internal workflow. For simpler tasks (e.g., geometry optimization), the workflow follows sequential logic. The agent prepares the necessary inputs—such as acquiring molecular structures and setting calculation parameters—in order, before finally invoking the MLatom code to execute the calculation. For more complex tasks (e.g., reaction energy calculation), the agent employs a reasoning process and workflow design similar to that of the orchestrating agent, involving the invocation of various tools to manage complexity.

***Interaction between Aitomia and the MLatom/HPC environment***

Aitomia employs a hybrid architecture that combines tool-calling for MLatom integration with a node-local execution model for HPC job management. In brief, the tool generates MLatom input files or Python scripts from predefined templates and saves them to the working directory so the user can later modify them. Then it uses the command line to run the input files or execute the Python script. In detail, the workflow operates as follows:

*Integration with MLatom*

The connection between Aitomia and the MLatom package is primarily achieved through function calls. We adopted this approach because LLMs are prone to hallucinations when generating code for functionalities with insufficient training data. Since computational stability is paramount, we chose not to delegate code generation to an LLM; instead, we generated an input file or Python script from a predefined template for a task.

The flexibility of our approach currently relies on a flexible combination of the single-task agents operating as described above. This design offers a viable strategy for working with code packages where the LLM may be under-trained or lack sufficient exposure. While this approach inevitably reduces some flexibility, we mitigate this by breaking tasks down into smaller, single-purpose units. One agent is responsible for one specific, granular task, and complex tasks are achieved by orchestrating these smaller units.

MLatom provides the computational backbone for Aitomia, with dozens of interfaces and support for a wide range of computational chemistry tasks. The integration with MLatom is not merely about utilizing machine learning methods; it is fundamentally about developability and extensibility.

*Integration with HPC clusters*

The Aitomia system adopts a node-local execution model. When Aitomia is submitted via the web interface, the scheduling system launches the main service script on an assigned compute node within the HPC cluster. All subsequent computational commands are executed from that node's local shell environment using system command calls. This approach ensures that computations directly utilize the node's allocated resources and the centrally maintained software stack accessible via the shared filesystem.

***Enhancing model robustness through retrieval-augmented generation (RAG)***

Although the LLM can, at the date of writing, often provide reasonable answers about computational chemistry and machine learning, it currently often hallucinates due to the vastness of the relevant knowledge. For example, it provides wrong answers about the capabilities of MLatom, which is at the core of our implementations. To address these issues and improve the model's performance, we also integrate RAG to enhance Aitomia's robustness as a lookup tool by providing access to external knowledge, thereby reducing hallucinations and improving factual accuracy (Figure 8). In Aitomia, external knowledge is first chunked and then represented as embeddings generated by the text-embedding-v3 model via the API provided by Aliyun.

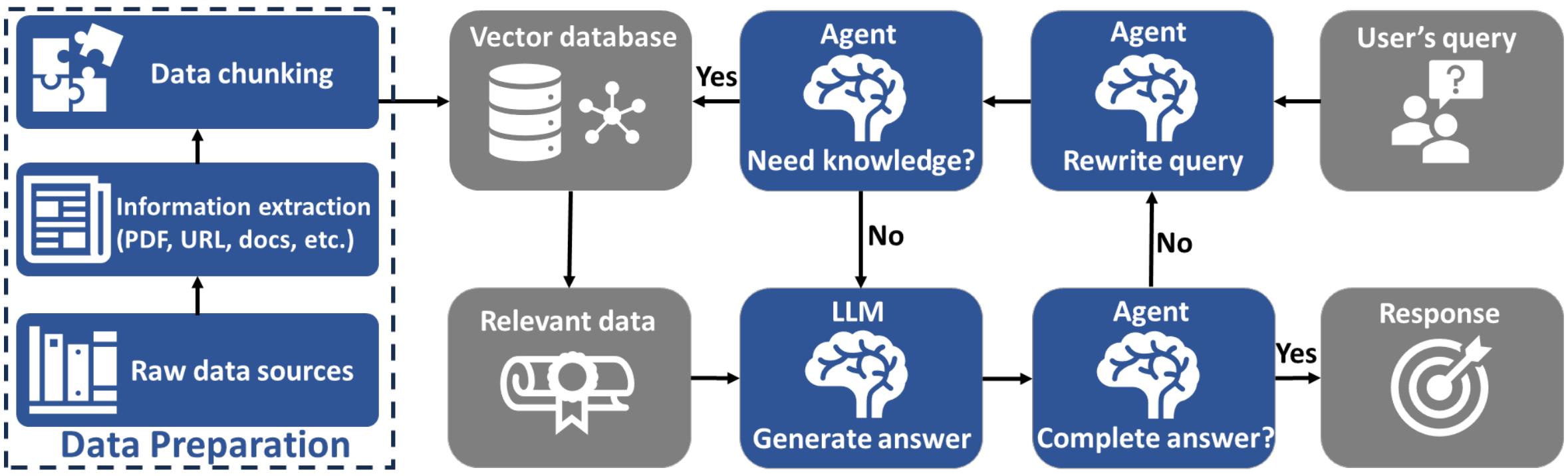

**Figure 8.** The Adaptive-RAG pipeline: retrieving external knowledge for enhanced generation.

Traditional single-step RAG, while technically the simplest and most feasible approach, is inaccurate for complex queries because it performs only a single retrieval step before generating an answer.[46, 47] The multi-step variant performs much better on complex queries but is inefficient with simple queries.

Adaptive-RAG presents a promising alternative,[48] which we have implemented in Aitomia. For straightforward queries, it can answer them directly without consulting the database; for simple queries, it checks the database only once to ensure efficiency; for complex queries, it can retrieve from the database multiple times until the queries are fully answered.[48]

When RAG is activated, Aitomia determines whether the question is related to the database's knowledge. If so, Aitomia retrieves information from the RAG database. Then LLM answers the user's query based on the retrieved information. After getting an intermediate answer, Aitomia judges whether it is enough to answer the query. If not, Aitomia updates the query and retrieves more information from the database. Finally, Aitomia makes a summary and provides a complete answer to the initial query.

By using this strategy, it retrieves only information relevant to the database, thereby reducing the number of retrievals and mitigating the model's tendency to hallucinate. Therefore, we integrate Adaptive-RAG to enhance the robustness of our model. The integration of Adaptive-RAG reduces the probability of model hallucinations and improves the model performance based on user needs.

### *Implementation details*

All the logic and APIs are implemented using JavaScript, HTML, the Python FastAPI library, and Uvicorn 0.34.0. In the RAG part, we use the text-embedding-v3 model as a vector database that generates embeddings of dimension 1024. We use `RecursiveCharacterTextSplitter` to make chunks, with `chunk_size=1000` and `chunk_overlap=200`. In total, 914 chunks are retrieved from the MLatom documentation. All AI agent connections and communications are coded using LangGraph (v.0.6.6). In all shown benchmarks, we use the Qwen3-Max-Preview LLM.

At the core of Aitomia's computational engine is MLatom, an open-source software package that integrates quantum chemical calculations with AI-enhanced atomistic simulations. MLatom can be easily installed locally via Python's package manager (`pip install mlatom`), enabling researchers to operate the entire platform within their own computing environment. MLatom is based on many third-party programs, as described in its documentation, including Gaussian,[7] PySCF,[49] Orca,[50, 51] MNDO,[52] xtb,[53] dftd4,[54] TorchANI,[55] mace,[56] sGDML,[57] PhysNet,[58] DeePMD-kit,[59] geomeTRIC,[60] and ASE[61]. Many of the state-of-the-art AI atomistic models are available as add-ons to MLatom via Aitomic.[62]

### *Platform availability*

The Aitomia intelligent assistant is publicly available on Aitomistic Lab@XMU[12] (https://atom.xmu.edu.cn, free for academic users) and Aitomistic Hub[11] (aitomistic.xyz, free registration) cloud computing platform.

### Author contributions

J.H. and H.N.: methodology, investigation, software, formal analysis, validation, writing – original draft, writing – review & editing, visualization, data curation. Y.H.: methodology, investigation, software, formal analysis, validation, writing – review & editing, visualization, figures, data curation. L.C.: software, visualization. Y.R.: formal analysis, validation, visualization, writing – review & editing. Y.C.: methodology, software, writing – review & editing. A.U.: conceptualization, methodology, writing – original draft, writing – review & editing, figures, supervision. P.O.D.: conceptualization, methodology, software, writing – original draft, writing – review & editing, visualization, resources, funding acquisition, supervision, project administration.

## Acknowledgments

P.O.D. acknowledges funding by the projects for International Senior Scientists (Project No.: W2531013) and for Outstanding Youth Scholars (Overseas, 2021) of the National Natural Science Foundation of China, and via the Lab project of the State Key Laboratory of Physical Chemistry of Solid Surfaces. A.U. acknowledges funding from the National Natural Science Foundation of China (No. W2433037) and Natural Science Foundation of Anhui Province (No. 2408085QA002). We also thank Aitomistic for software development and for integrating Aitomia into Aitomistic Lab, Aitomistic Lab@XM[12] (https://atom.xmu.edu.cn), and Hub (aitomistic.xyz), making it publicly available.

## Declaration of interests

The authors declare no competing interests.